\documentclass[prd,showpacs,preprintnumbers,amsmath,amssymb]{revtex4}

\usepackage{graphicx}
\usepackage{dcolumn}
\usepackage{bm}

\begin{document}

\preprint{hep-th/0005078}

\title{Schwinger Pair Production via Instantons in Strong Electric Fields}

\author{Sang Pyo Kim}\email{sangkim@ks.kunsan.ac.kr; spkim@phys.ualberta.ca}
\affiliation{Department of Physics,
Kunsan National University, Kunsan 573-701, Korea\\
and Theoretical Physics Institute, Department of Physics,
University of Alberta, Edmonton, Alberta, Canada T6G 2J1}

\author{Don N. Page}\email{don@phys.ualberta.ca}
\affiliation{CIAR Cosmology Program, Theoretical Physics
Institute, Department of Physics, University of Alberta, Edmonton,
Alberta, Canada T6G 2J1}

\date{\today}
\begin{abstract}
In the space-dependent gauge, each mode of the Klein-Gordon
equation in a strong electric field takes the form of a
time-independent Schr\"{o}dinger equation with a potential
barrier. We propose that the single- and multi-instantons of
quantum tunnelling may be related with the single- and multi-pair
production of bosons and the relative probability for the no-pair
production is determined by the total tunnelling probability via
instantons. In the case of a static uniform electric field, the
instanton interpretation recovers exactly the well-known pair
production rate for bosons and when the Pauli blocking is taken
into account, it gives the correct fermion production rate. The
instanton is used to calculate the pair-production rate even in an
inhomogeneous electric field.  Furthermore, the instanton
interpretation confirms the fact that bosons and fermions cannot
be produced by a static magnetic field only.
\end{abstract}
\pacs{PACS number(s): 11.15.Kc, 12.20.Ds, 04.60.+v}

\maketitle

\section{Introduction}

Strong electromagnetic fields lead to two physically important
phenomena: the pair-production and vacuum polarization. A strong
electric field makes the quantum electrodynamics (QED) vacuum
unstable which decays by emitting significantly boson or fermion
pairs \cite{sau,hei,sch}. The vacuum fluctuations of an external
electromagnetic field also result in an effective action of the
nonlinear Maxwell equations \cite{hei,sch,wei}.  As its long
history, there have been developed many different methods such as
the proper time method \cite{sch,dew}, canonical method
\cite{dew}, etc., to derive the QED effective action in external
electromagnetic fields. Also there have been applications to
various physical problems \cite{gre}. The proper time method by
Schwinger \cite{sch} and DeWitt \cite{dew} has widely been
employed to compute the effective action. The real part of the
effective action leads to the vacuum polarization and the
imaginary part to the pair-production. Though that method is
conceptually well-defined and technically rigorous, it is
sometimes difficult to apply the method to some concrete physical
problems such as inhomogeneous electromagnetic fields and others.
On the other hand, the canonical method \cite{dew} proves quite
efficient in calculating the pair-production rate of bosons and
fermions by static or time-dependent uniform electric fields in
many physical contexts.

In canonical approach the most frequently used gauge for the
electromagnetic potential is the time-dependent gauge. In that
gauge the Klein-Gordon equation for bosons or the Dirac equation
for fermions in a uniform electric field, when appropriately
mode-decomposed, takes the form of time-dependent Schr\"{o}dinger
equations. Now the pair-production by the external electric field
is analogous to the particle production by a time-dependent metric
of a curved spacetime \cite{par,par2,par3}. In both problems one
imposes the same boundary condition that an incident, positive
frequency component in the past infinity is scattered by a
potential barrier into a superposition of positive and negative
frequency components in the future infinity. It is the complex
conjugate of the boundary condition for scattering problems in
quantum mechanics. The coefficients determine the Bogoliubov
transformation and, in particular, the coefficient of the negative
frequency component gives the number created bosons or fermions
per mode. The pair-production rates were calculated for
time-varying electric fields
\cite{bre,nar,pop,pop2,cor,bal,klu,dun}. Using both canonical and
path integral methods, the pair-production in a uniform electric
field was studied in the time-dependent gauge \cite{pad,pad2} and
in Rindler coordinates \cite{gab}. The pair-production was also
studied for a uniform electric field confined to a finite region,
an inhomogeneous field \cite{wan,mar,mar2}

A shortcoming of the time-dependent gauge is that except for
uniform fields, the gauge potential and thereby the Klein-Gordon
equation involve both the space and time coordinates at the same
time. So it is technically difficult to apply the Bogoliubov
transformation for inhomogeneous fields. On the other hand, in the
space-dependent (Coulomb) gauge for a static electric field, each
mode of the Klein-Gordon equation for bosons or the Dirac equation
for fermions takes the form of a time-independent Schr\"{o}dinger
equation for quantum tunnelling through a potential barrier. In
that space-dependent gauge there is no direct interpretation of
wave components in terms of positive and negative frequencies.
However, in the case of the static uniform electric field, Brezin
and Itzykson explained the dominant contribution to the
pair-production rate by quantum tunnelling through the potential
barrier \cite{bre,itz}, and Casher {\it et al} rederived the
Schwinger's pair-production rate by semiclassical tunnelling
calculation \cite{neu,neu2,neu3}. Nikishov found the
pair-production rate in scattering matrix formalism for a uniform
field and an inhomogeneous field of Sauter type gauge potential
\cite{nik}. Hansen and Ravndal  showed that the transmission
probability through the barrier of a uniform electric field gives
the probability for pair-production for bosons and fermions
\cite{han}, solving the Klein paradox \cite{kle,hun,sau}. Also
Padmanabhan \cite{pad2} suggested that the reflection probability
of the scattering problem gives the correct relative probability
for the vacuum-to-vacuum transition for bosons. The role of
tunnelling solutions for pair-production was also noticed in
Refs.\cite{pop,ste,bro,bro2,pare}.

The purpose of this paper is to interpret and derive the boson or
fermion pair-production rate by strong static uniform or
inhomogeneous electric fields in terms of instantons through
potential barriers in the space-dependent gauge in any spacetime
dimensions. This formula in terms of the instanton action may
provide a simple way to estimate the pair-production rate by
inhomogeneous electric fields, for instance, from charged black
holes, neutron stars, or astrophysical objects \cite{ruf,ruf2}.
For these static inhomogeneous fields it is easier to apply the
space-dependent gauge than the time-dependent gauge. We propose
that the single- and multi-instantons for quantum tunnelling
determine somehow the single- and multi-pair production. In
particular, we show that all the contributions from
multi-instantons and anti-instantons yield exactly the total
tunnelling probability for the static uniform electric field, and
thereby determine the relative vacuum-to-vacuum transition and the
boson pair-production rate. We further show that the instanton
interpretation together with the Pauli blocking gives correctly
the fermion production rate by the static uniform electric field.
Using the formula in terms of the instanton action, we find the
pair-production rates for bosons and fermions which are
asymptotically valid for extremely strong electric fields. Also
the pair-production rates for bosons and fermions by a static
inhomogeneous electric field are calculated using WKB (adiabatic)
approximation for the instantons. Finally we show that according
to the instanton interpretation a static localized magnetic field
does not lead to any pair-production, confirming the result from
the proper time method.

The organization of this paper is as follows. In Sec. II we show
that the tunnelling probability by instantons gives correctly the
pair-production rates for bosons and fermions by a static uniform
electric field. We calculate the pair-production rates in any
spacetime dimensions and find their asymptotic form for extremely
strong field and compare them with those from other methods. In
Sec. III we extend the instanton interpretation of pair-production
to an inhomogeneous electric field and find the pair-production
rates in terms of the instanton action. In Sec. IV we apply the
idea to a static magnetic field to show that any pair of boson or
fermion are not produced. This resolves some of the puzzling issue
in the canonical method on the pair-production by a static
localized magnetic field.

\section{Uniform Electric Field}

We consider a charged boson in a static uniform electric field in
a $(d+1)$-dimensional Minkowski spacetime. It satisfies the
Klein-Gordon equation (in units of $\hbar = c =1$)
\begin{equation}
\Biggl[\eta^{\mu \nu} \Biggl(\frac{\partial}{\partial x^{\mu}} + i
q A_{\mu} \Biggr) \Biggl(\frac{\partial}{\partial x^{\nu}} + i q
A_{\nu} \Biggr)  + m^2 \Biggr] \Phi (t, {\bf x}) = 0, \label{kg
eq}
\end{equation}
where $q$ is the charge and $m$ the mass of the boson. In the
space-dependent (Coulomb) gauge, the vector potential for the
uniform electric field in the $x_{\parallel}$-direction is given
by
\begin{equation}
A_{\mu} (t, {\bf x}) = ( - E_0 x_{\parallel}, 0, \cdots, 0).
\end{equation}
Each Fourier-mode of the boson field
\begin{equation}
\Phi (t, {\bf x}) = e^{i ( {\bf k}_{\perp} \cdot {\bf x}_{\perp} -
\omega t)} \phi_{\omega, {\bf k}_{\perp}} (x_{\parallel}),
\end{equation}
satisfies the one-dimensional equation
\begin{equation}
\Biggl[- \frac{1}{2} \frac{d^2}{dx_{\parallel}^2} -
\frac{1}{2}\Biggl( \omega + q E_0 x_{\parallel} \Biggr)^2 \Biggr]
\phi_{\omega, {\bf k}_{\perp}} (x_{\parallel}) = - \frac{1}{2}(m^2
+ {\bf k}_{\perp}^2) \phi_{\omega, {\bf k}_{\perp}}
(x_{\parallel}). \label{mod eq}
\end{equation}

Now one may interpret Eq. (\ref{mod eq}) as a Schr\"{o}dinger-like
equation for a unit mass moving in the inverted harmonic potential
with the center at $x_{\parallel, c} = - \omega/(qE_0)$ and the
energy $\epsilon = - (m^2 + {\bf k}_{\perp}^2)/2$. As the energy
is negative $(\epsilon < 0)$, Eq. (\ref{mod eq}) indeed describes
a tunnelling problem for all transverse momenta ${\bf k}_{\perp}$.
The wave function describing the tunnelling process is given by
the complex parabolic cylindrical function \cite{abr}
\begin{equation}
\phi_{\omega, {\bf k}_{\perp}} (\xi) = c E(a_{{\bf k}_{\perp}},
\xi),
\end{equation}
where $c$ is a complex number, and
\begin{equation}
\xi = \sqrt{\frac{2}{q E_0}} ( \omega + q E_0 x_{\parallel}),
\quad a_{{\bf k}_{\perp}} = \frac{m^2 + {\bf k}_{\perp}^2}{2 q
E_0}.
\end{equation}
It has the asymptotic forms in two regimes
\begin{eqnarray}
\phi_{\omega, {\bf k}_{\perp}} (\xi) &=& A \varphi_{\omega, {\bf
k}_{\perp}} (\xi) - B \varphi^*_{\omega, {\bf k}_{\perp}} (\xi),
\quad (\xi  \ll -2 \sqrt{a_{{\bf k}_{\perp}}}),
 \nonumber\\
\phi_{\omega, {\bf k}_{\perp}} (\xi) &=& C \varphi^*_{\omega, {\bf
k}_{\perp}} (\xi), \quad (\xi  \gg 2 \sqrt{a_{{\bf k}_{\perp}}}),
\label{asym}
\end{eqnarray}
where
\begin{equation}
\varphi_{\omega, {\bf k}_{\perp}} (\xi) = \sqrt{\frac{2}{|\xi|}}
e^{- \frac{i}{4} \xi^2 }.
\end{equation}
Here the coefficients are given by
\begin{equation}
A = i c \sqrt{1 + e^{2 \pi a_{{\bf k}_{\perp}}}}, \quad B = - i c
e^{\pi a_{{\bf k}_{\perp}}}, \quad C= c.
\end{equation}

In the region $\xi \ll - 2 \sqrt{a_{{\bf k}_{\perp}}}$, the
components $\varphi_{\omega, {\bf k}_{\perp}} e^{- i \omega t}$
describes an incoming particle, from and $\varphi^*_{\omega, {\bf
k}_{\perp}} e^{- i \omega t}$ an outgoing particle, to $\xi= -
\infty$, whereas in the region $\xi \gg 2 \sqrt{a_{{\bf
k}_{\perp}}}$ the component $\varphi^*_{\omega, {\bf k}_{\perp}}
e^{- i \omega t}$ describes an incoming anti-particle from $\xi =
+ \infty$.  Hansen and Ravndal showed that the transmission
probability $\vert C/ A \vert^2$ gives the probability for
one-pair production \cite{han}. Also Padmanabhan suggested that
the reflection probability $\vert B/ A \vert^2$ gives the relative
probability for the vacuum-to-vacuum transition \cite{pad,pad2}.
His interpretation implies that $\varphi_{\omega, {\bf k}_{\perp}}
(- \infty) e^{- i \omega t}$ and $\varphi^*_{\omega, {\bf
k}_{\perp}} (- \infty) e^{- i \omega t}$ correspond to the
incoming and outgoing vacuum state, respectively. Extending their
arguments to any static field, we further propose that the single-
and multi-pair production of bosons are related to the single- and
multi-instantons of potential barrier in such a way that the
tunnelling probability $P^{\rm t}$ gives the probability for the
pair-production and therefore the relative probability for the
vacuum-to-vacuum transition is given by the probability for the
no-pair production $P^{\rm n-p} = 1 - P^{\rm t}$. Here and from
now on we restrict the {\it tunnelling probability} to the
transmission probability through potential barrier but exclude any
nonzero transmission probability above a potential barrier or a
potential well. Further we shall assume that the tunnelling
probability is accurately given by the instanton action or with
its higher corrections.

To see how the instanton interpretation works for the uniform
electric field, we calculate the tunnelling probability from the
asymptotic form (\ref{asym}) and compare it with the result from
the instanton calculation. As the negative energy for all momenta
${\bf k}_{\perp}$ is below the potential barrier in Eq. (\ref{mod
eq}), the tunnelling probability is given by the transmission
probability
\begin{equation}
P^{\rm b.t}_{{\bf k}_{\perp}} = \Biggl|\frac{C}{A} \Biggr|^2 =
\frac{1}{e^{2 \pi a_{{\bf k}_{\perp}}} + 1}. \label{tun}
\end{equation}
Likewise, the probability for the no-pair production, {\it i.e.},
the vacuum-to-vacuum transition, given by the reflection
probability
\begin{equation}
P^{\rm b.n-p}_{{\bf k}_{\perp}} = 1 - P^{\rm b.t}_{{\bf
k}_{\perp}} = \frac{1}{1 + e^{ - 2 \pi a_{{\bf k}_{\perp}}}} =
\Biggl|\frac{B}{A} \Biggr|^2, \label{ref}
\end{equation}
as a consequence of the flux conservation. Hence, what is needed
in finding the probability for the no-pair production
(vacuum-to-vacuum transition) even in a general electric field is
the corresponding total tunnelling probability via the single- and
multi-instantons.

Now let us interpret the tunnelling probability (\ref{tun}) in
terms of multi-instantons and anti-instantons of tunnelling
process. In instanton physics \cite{col}, the leading contribution
to the tunnelling probability
\begin{equation}
P^{\rm t}_{{\bf k}_{\perp}} = e^{- 2 S_{{\bf k}_{\perp}}},
\end{equation}
is determined by the single-instanton action
\begin{equation}
S_{{\bf k}_{\perp}} = \int_{x_-}^{x^+} dx_{\parallel} \sqrt{m^2 +
{\bf k}_{\perp}^2 - \Bigl( \omega + qE_0 x_{\parallel} \Bigr)^2 }
= \pi  a_{{\bf k}_{\perp}}, \label{inst}
\end{equation}
where $x_{\pm} = \pm \sqrt{m^2 + {\bf k}_{\perp}^2} - \omega$ are
the classical turning points. We propose that the single-instanton
and multi-instantons may be related in a certain way with one-pair
and multi-pair production, whereas multi-anti-instantons with the
annihilation of created boson pairs. As there is no limitation
from the Pauli blocking for the multi-pair production of bosons,
the correct total tunnelling probability should take into account
both multi-instantons and anti-instantons
\begin{equation}
P^{\rm b.t}_{{\bf k}_{\perp}} = \sum_{n = 1}^{\infty} (-1)^{n+1}
e^{- 2n S_{{\bf k}_{\perp}}} = \frac{1}{ e^{2S_{{\bf k}_{\perp}}}
+ 1},
\end{equation}
where instantons contribute positively and anti-instantons
negatively. Similarly, the relative probability for the no-pair
production (vacuum-to-vacuum transition) is given by
\begin{equation}
P^{\rm b.n-p}_{{\bf k}_{\perp}} =  \sum_{n = 0}^{\infty} (-1)^{n}
e^{- 2n S_{{\bf k}_{\perp}}} = \frac{1}{1 + e^{- 2S_{{\bf
k}_{\perp}}}},
\end{equation}
These results agree with Eqs. (\ref{tun}) and (\ref{ref}). The
physical interpretation of the alternating signs is that only the
instantons of even repeated periodic motions in the inverted
potential contribute positively (creating pairs) to the tunnelling
probability, whereas the anti-instantons of odd repeated periodic
motions contribute negatively (annihilating created-pairs) to the
tunnelling probability.

The vacuum means the absence of any particle for possible physical
states. So the vacuum-to-vacuum transition, {\it i.e.}, the vacuum
persistence, is the total relative probability for the no-pair
production:
\begin{equation}
\vert \langle 0, {\rm out} \vert 0, {\rm in} \rangle \vert^2 =
\prod_{\rm all~ states} P^{\rm b.n-p}_{{\bf k}_{\perp}} = \exp
\Biggl[ - \sum_{\rm all~states} \ln (1 + e^{- 2S_{{\bf
k}_{\perp}}}) \Biggr].
\end{equation}
On the other hand, the vacuum-to-vacuum transition is given by the
imaginary part of the effective action for boson
\begin{equation}
\vert \langle 0, {\rm out} \vert 0, {\rm in} \rangle \vert^2 =
\exp \Biggl[ - 2 V T {\rm Im} {\cal L}^{\rm b}_{\rm eff.} \Biggr],
\end{equation}
where $V$ and $T$ are the relevant volume and the duration of
time. Therefore, the pair-production rate per unit time per unit
volume is twice of the imaginary part of the effective action:
\begin{equation}
w^{\rm b} =  2{\rm Im} {\cal L}^{\rm b}_{\rm eff.} = \frac{1}{V T}
 \sum_{\rm all~states} \ln (1 + e^{- 2S_{{\bf
k}_{\perp}}}). \label{b-p}
\end{equation}
Then the pair-production rate for bosons is explicitly given by
\begin{eqnarray}
w^{\rm b} &=& \frac{(2s+1)V_{\perp}}{V}  \int \frac{d \omega d{\bf
k}_{\perp}^{d-1}}{(2\pi)^d} \sum_{n = 1}^{\infty}
\frac{(-1)^{n+1}}{n} e^{- \frac{\pi n}{q E_0} {\bf k}_{\perp}^2}
e^{- \frac{\pi m^2}{q E_0} n} \nonumber\\ &=&
\frac{(2s+1)}{(2\pi)^d} \sum_{n = 1}^{\infty}
(-1)^{n+1}\Biggl(\frac{q E_0}{n} \Biggr)^{(d+1)/2} e^{- \frac{\pi
m^2}{q E_0} n}, \label{im eff}
\end{eqnarray}
where $s$ is the spin of the boson. Here we used $\int d \omega =
(qE_0) V_{\parallel}$, where $V_{\parallel}$ is the longitudinal
extension of the field, and $V = V_{\perp} V_{\parallel}$,
$V_{\perp}$ being the transverse volume \cite{nik}. It should be
noted that Eq. (\ref{im eff}) recovers the standard result for the
boson pair-production in any dimension in Ref. \cite{gus}.

The fermion pair-production can be understood similarly. The
created fermion pair blocks the multi-pair production. So the
total tunnelling probability for the fermion pair-production per
each mode is just
\begin{equation}
P^{\rm f.t}_{{\bf k}_{\perp}} = e^{-2S_{{\bf k}_{\perp}}}.
\end{equation}
Therefore, the relative probability for the no-pair production of
fermions is now given by
\begin{equation}
P^{\rm f.n-p}_{{\bf k}_{\perp}} =1 -  e^{-2S_{{\bf k}_{\perp}}}.
\end{equation}
Finally, the fermion pair-production rate per unit time per unit
volume is found to be
\begin{equation}
w^{\rm f}  = 2 {\rm Im} {\cal L}^{\rm f}_{\rm eff.} = -
\frac{1}{VT} \sum_{\rm all~states} \ln (1 - e^{- 2S_{{\bf
k}_{\perp}}}), \label{f-p}
\end{equation}
and takes the form
\begin{eqnarray}
w^{\rm f} &=& \frac{(2s+1)V_{\perp}}{V}  \int \frac{d \omega d{\bf
k}_{\perp}^{d-1}}{(2\pi)^d} \sum_{n = 1}^{\infty} \frac{1}{n} e^{-
\frac{\pi n}{q E_0} {\bf k}_{\perp}^2} e^{- \frac{\pi m^2}{q E_0}
n} \nonumber\\ &=& \frac{(2s+1)}{(2\pi)^d} \sum_{n = 1}^{\infty}
\Biggl(\frac{q E_0}{n} \Biggr)^{(d+1)/2} e^{- \frac{\pi m^2}{q
E_0} n}. \label{f-im}
\end{eqnarray}
Also Eq. (\ref{f-im}) recovers the standard result for the fermion
pair-production in Ref. \cite{gus}.

Though the production rate (\ref{im eff}) for bosons and
(\ref{f-im}) for fermions are well defined for all electric
fields, the series converge strongly for weak electric fields
because all higher terms are exponentially suppressed. But for
extremely strong electric fields the exponential terms approach to
unity and the series are approximated by the Riemann eta function
$\eta (2)$ for bosons and the Riemann zeta function $\zeta (2)$
for fermions. So, for strong electric fields, instead of using a
special resummation of the series, we adopt directly the
pair-production formula (\ref{b-p}) and (\ref{f-p}) and evaluate
properly the integrals suitable for strong fields. In four
dimensions $(d = 3)$, the boson pair-production rate (\ref{b-p})
becomes
\begin{eqnarray}
w^{\rm b} &=& \frac{(2s+1)}{(2 \pi)^3} (qE_0) \int_0^{\infty} (2
\pi) d k_{\perp} k_{\perp} \ln \Biggl(1 + e^{- \frac{\pi(m^2 +
k_{\perp}^2 )}{qE_0}} \Biggr) \nonumber\\ &=& \frac{(2s+1)}{(2
\pi)^3} (qE_0)^2 \Biggl\{\int_0^{\infty} dy \ln (1 + e^{-y}) -
\int_0^{\pi m^2/qE_0} dy \ln ( 1 + e^{-y}) \Biggr\},
\end{eqnarray}
where
\begin{equation}
y = \frac{\pi}{qE_0} k_{\perp}^2 + \frac{\pi m^2}{q E_0}.
\end{equation}
Using the integral \cite{pbm}
\begin{equation}
\int_0^{\infty} dy \ln (1 + e^{-y}) = \frac{\pi^2}{12},
\end{equation}
and expanding the exponential and then the logarithmic function to
any desired order
\begin{equation}
\ln ( 1 + e^{-y}) = \ln 2 - \frac{1}{2} y + \frac{1}{8} y^2 +
\frac{1}{96} y^4 + {\cal O} (y^5),
\end{equation}
we obtain the pair-production rate
\begin{eqnarray}
w^{\rm b} &=& \frac{(2s+1)}{(2 \pi)^3} \Biggl\{ \frac{\pi^2}{12}
(qE_0)^2 - (\ln 2) \pi m^2 q E_0 + \frac{1}{4} (\pi m^2)^2
\nonumber\\~~~~&-& \frac{1}{24} \frac{(\pi m^2)^3}{qE_0} -
\frac{1}{480} \frac{(\pi m^2)^5}{(qE_0)^3}  + {\cal O}
\Biggl(\frac{(\pi m^2)^6}{(qE_0)^4} \Biggr) \Biggr\}. \label{s
b-p}
\end{eqnarray}
Similarly, the fermion pair-production rate (\ref{f-p}) for strong
fields takes the form
\begin{eqnarray}
w^{\rm f} &=& - \frac{(2s+1)}{(2 \pi)^3} (qE_0) \int_0^{\infty} (2
\pi) d k_{\perp} k_{\perp} \ln \Biggl(1 - e^{- \frac{\pi(m^2 +
k_{\perp}^2 )}{qE_0}} \Biggr) \nonumber\\ &=& \frac{(2s+1)}{(2
\pi)^3} (qE_0)^2 \Biggl\{\int_0^{\infty} dy \ln (1 - e^{-y}) -
\int_0^{\pi m^2/qE_0} dy \ln ( 1 - e^{-y}) \Biggr\}.
\end{eqnarray}
Using the integral \cite{pbm}
\begin{equation}
\int_0^{\infty} dy \ln (1 - e^{-y}) = - \frac{\pi^2}{6},
\end{equation}
and expanding the exponential function and then the logarithmic
function
\begin{equation}
\ln ( 1 - e^{-y}) = \ln y - \frac{1}{2} y + \frac{1}{24} y^2 +
\frac{11}{720} y^4 + {\cal O} (y^5),
\end{equation}
we finally obtain the fermion pair-production rate
\begin{eqnarray}
w^{\rm f} &=& \frac{(2s+1)}{(2 \pi)^3} \Biggl\{ \frac{\pi^2}{6}
(qE_0)^2 - \pi m^2 q E_0 \Biggl(\ln \Biggl(\frac{qE_0}{\pi m^2}
\Biggr) + 1 \Biggr) - \frac{1}{4} (\pi m^2)^2 \nonumber\\~~~~&+&
\frac{1}{72} \frac{(\pi m^2)^3}{qE_0} + \frac{11}{3600} \frac{(\pi
m^2)^5}{(qE_0)^3}  + {\cal O} \Biggl(\frac{(\pi m^2)^6}{(qE_0)^4}
\Biggr) \Biggr\}. \label{s f-p}
\end{eqnarray}
The fermion pair-production rate (\ref{s f-p}) for strong electric
fields confirms the result obtained from different methods in
Refs. \cite{dit,hey,sol}.

A comment is in order. The Schwinger pair-production by a static
uniform electric field is an ideal calculation in which one
neglects the pair-production due to the interactions of the
created pairs with the electric field background and among the
created pairs. For instance, a single-pair can produce another
pair through the interaction with the electric field, whose rate
is proportional to $(qE_0/m^2)^2 (q/m)^2$ \cite{jau} and can be
larger than the multi-pair production rate, $e^{- (\pi m^2
n)/(qE_0)} (qE_0)^2/n^2$, from multi-instantons for all
sufficiently large $n$ even for an extremely strong electric field
$E_0$. However, we shall not consider this complicated real
situation but rather focus on the ideal calculation without the
back-reaction of produced pairs.

\section{Inhomogeneous Electric Fields}

We now consider the pair-production by a static inhomogeneous
electric field. Without loss of generality, the electric field is
assumed to be localized in the $x_{\parallel}$-direction and to
have the gauge potential
\begin{equation}
A_{\mu} (t, {\bf x}) = ( A_0 (x_{\parallel}), 0, \cdots, 0),
\end{equation}
where $E (x_{\parallel}) = - d A_0
(x_{\parallel})/dx_{\parallel}$. We restrict only to the case
where all produced particles $(q > 0)$ and antiparticles reach the
asymptotic regions $x = + \infty$ and $x = - \infty$,
respectively, without being bounded by the electric field. This
requires that $q A_0 (- \infty) - q A_0 (+ \infty) \geq 2m$. The
mode-decomposed Klein-Gordon equation then takes the form
\begin{equation}
\Biggl[- \frac{1}{2} \frac{d^2}{dx_{\parallel}^2} -
\frac{1}{2}\Biggl(\omega - q A_0 ( x_{\parallel}) \Biggr)^2
\Biggr] \phi_{\omega, {\bf k}_{\perp}} (x_{\parallel}) = -
\frac{1}{2}(m^2 + {\bf k}_{\perp}^2) \phi_{\omega, {\bf
k}_{\perp}} (x_{\parallel}). \label{inh mod eq}
\end{equation}
We can still interpret Eq. (\ref{inh mod eq}) as a one-dimensional
quantum system of a unit mass with the potential $- (\omega - q
A_0 ( x_{\parallel}))^2/2$ and the energy $\epsilon = - (m^2 +
{\bf k}_{\perp}^2)/2$. In the WKB (adiabatic) approximation the
asymptotic form for the tunnelling probability for each mode ${\bf
k}_{\perp}$ is given by \cite{fro,fro2,ben}
\begin{equation}
P^{\rm b.t}_{{\bf k}_{\perp}} = \frac{1}{e^{2 S_{{\bf k}_{\perp}}}
+1},
\end{equation}
where
\begin{equation}
S_{{\bf k}_{\perp}} = \sum_{n=0}^{\infty} S^{(2n)}_{{\bf
k}_{\perp}}.
\end{equation}
Here the leading contribution to $S_{{\bf k}_{\perp}}$ is given by
the instanton action
\begin{equation}
S^{(0)}_{{\bf k}_{\perp}} =  \int_{x_-}^{x_+} dx_{\parallel}
\Biggl[ Q_{{\bf k}_{\perp}}(x) \Biggr]^{1/2}, \label{inh inst}
\end{equation}
and the next-to-leading term by
\begin{equation}
S^{(2)}_{{\bf k}_{\perp}} = \int_{x_-}^{x_+} dx_{\parallel}
\Biggl[\frac{1}{8}\frac{Q_{{\bf k}_{\perp}}''(x)}{Q^{3/2}_{{\bf
k}_{\perp}}(x)} - \frac{5}{32}\frac{Q^{'2}_{{\bf
k}_{\perp}}(x)}{Q^{5/2}_{{\bf k}_{\perp}}(x)} \Biggr],
\end{equation}
where
\begin{equation}
Q_{{\bf k}_{\perp}}(x) = m^2 + {\bf k}_{\perp}^2 - \Bigl( \omega -
q A_0 ( x_{\parallel}) \Bigr)^2.
\end{equation}
Hence the relative probability for the no-pair production
(vacuum-to-vacuum transition) of bosons is given by
\begin{equation}
P^{\rm b.n-p}_{{\bf k}_{\perp}} = \frac{1}{1+ e^{- 2 S_{{\bf
k}_{\perp}}}},
\end{equation}
and for fermions by
\begin{equation}
P^{\rm f.n-p}_{{\bf k}_{\perp}} = 1 - e^{- 2 S_{{\bf k}_{\perp}}}.
\end{equation}

A few comments are in order. First, if the electric field extends
over all the space as in the uniform field case and has the
potential $\vert A_0 (\pm \infty) \vert = \infty$, then the
potential barrier decreases indefinitely at both $\pm \infty$.
Therefore, there are always instantons for all ${\bf k}_{\perp}$.
Second, if the electric field is localized or has finite values of
the potential at $\pm \infty$, then pairs are produced only when
$\omega - q A_0 (+\infty) \geq m$ and $\omega - q A_0 (- \infty)
\leq - m$. So there is a change of the sign of $\omega - q A_0
(x_{\parallel})$ implying a potential barrier. Thus only those
modes belonging to $ \vert {\bf k}_{\perp} \vert \leq k_{\perp,
{\rm max}}$ have finite instantons and lead to pair-production,
where the upper limit is given by the minimum of two asymptotic
values
\begin{equation}
{k}_{\perp, {\rm max}}^2 = {\rm Min} \Bigl\{ \Bigl(\omega - q A_0
(+ \infty) \Bigr)^2 - m^2, ~\Bigl(\omega - q A_0 (- \infty)
\Bigr)^2 - m^2 \Bigr\}.
\end{equation}

In the inhomogeneous electric field, we obtain the boson
pair-production rate per unit time per unit volume
\begin{eqnarray}
w^{\rm b} &=& 2 {\rm Im} {\cal L}^{\rm b}_{\rm eff.}\nonumber\\
&=& \frac{(2s +1)}{VT} \sum_{\rm all ~allowed ~states} \ln \Bigl(1
+ e^{- 2 S_{{\bf k}_{\perp}}} \Bigr) \nonumber\\ &=& \frac{(2s
+1)}{(2 \pi)^d V_{\parallel}} \frac{(d-1)
\pi^{(d-1)/2}}{\Gamma(\frac{d+1}{2})} \sum_{n = 1}^{\infty}
\frac{(-1)^{n+1}}{n} \int^{ qA_0(- \infty) - m}_{qA_0 (+ \infty) +
m} d\omega \int_{0}^{k_{\perp, {\rm max}}} dk_{\perp}
k_{\perp}^{d-2} e^{- 2 n S_{{\bf k}_{\perp}}}, \label{inh bos
rate}
\end{eqnarray}
and the fermion pair-production rate
\begin{eqnarray}
w^{\rm f} &=& 2 {\rm Im} {\cal L}^{\rm f}_{\rm eff.} \nonumber\\
&=& - \frac{(2s +1)}{2 VT} \sum_{\rm all~ allowed ~states} \ln
\Bigl(1 - e^{- 2 S_{{\bf k}_{\perp}}} \Bigr) \nonumber\\ &=&
\frac{(2s +1)}{2 (2 \pi)^d V_{\parallel}} \frac{(d-1)
\pi^{(d-1)/2}}{\Gamma(\frac{d+1}{2})} \sum_{n = 1}^{\infty}
\frac{1}{n} \int^{qA_0(- \infty) - m}_{qA_0 (+ \infty) + m}
d\omega \int_{0}^{k_{\perp, {\rm max}}} dk_{\perp} k_{\perp}^{d-2}
e^{- 2 n S_{{\bf k}_{\perp}}}. \label{inh fer rate}
\end{eqnarray}

As an exactly solvable model we consider a localized electric
field $E(x_{\parallel}) = E_0 {\rm sech}^2(x_{\parallel}/L)$ with
the Sauter type gauge potential \cite{sau,nik}
\begin{equation}
A_0 (x_{\parallel}) = - E_0 L \tanh (\frac{x_{\parallel}}{L}).
\label{sauter}
\end{equation}
In the limit of $L \rightarrow \infty$ the gauge potential
(\ref{sauter}) reduces to the uniform electric field in Sec. II.
Since the gauge potential (\ref{sauter}) is a more general case
including the uniform field as a special case, it is worthy to
apply the instanton interpretation to the pair-production and
compare the result with the exact one. Bosons gain an additional
contribution to momenta from the acceleration by the localized
electric field and have asymptotic values at $x_{\parallel}
\rightarrow \pm \infty$:
\begin{equation}
k_{\parallel}^2 (\infty) = (qE_0 L +\omega)^2 - m^2 - {\bf
k}^2_{\perp}, \quad k_{\parallel}^2 (- \infty) = (qE_0 L -
\omega)^2 - m^2 - {\bf k}^2_{\perp}.
\end{equation}

In the large $L$ limit the instanton action (\ref{inh inst}) is
given by
\begin{equation}
S_{{\bf k}_{\perp}} = \pi \frac{m^2 + {\bf k}^2_{\perp}}{2qE_0}
\Biggl[1 + \frac{\omega^2}{(qE_0 L)^2} +  \frac{m^2 + {\bf
k}^2_{\perp}}{4(qE_0L)^2} + {\cal O} (\frac{1}{L^4}) \Biggr].
\label{s inst}
\end{equation}
The exact wave function describing the tunnelling process is found
\begin{equation}
\phi_{\omega, {\bf k}_{\perp}} (x_{\parallel}) = C e^{- \mu
\frac{x_{\parallel}}{L}} {\rm sech}^{\nu}
(\frac{x_{\parallel}}{L}) F (\alpha, \beta; \gamma; \zeta),
\label{s wave}
\end{equation}
where $F$ is the hypergeometric function and
\begin{eqnarray}
\mu &=&  - i \frac{L}{2}\Bigl( k_{\parallel} (\infty)+
k_{\parallel} (- \infty) \Bigr), \quad \nu =  - i
\frac{L}{2}\Bigl( k_{\parallel} (\infty) - k_{\parallel} (-
\infty) \Bigr), \nonumber\\ \alpha &=& \nu + \frac{1}{2} + i
\sqrt{ (q E_0 L^2)^2 - \frac{1}{4}}, \quad \beta = \nu +
\frac{1}{2} - i \sqrt{(q E_0 L^2)^2 - \frac{1}{4}}, \nonumber\\
\gamma &=& 1 - i L k_{\parallel} (\infty), \quad \zeta =
\frac{1}{2} \Bigl(1 - \tanh(\frac{x_{\parallel}}{L}) \Bigr).
\end{eqnarray}
In the limit of $x_{\parallel} \gg L$, Eq. (\ref{s wave}) has the
asymptotic form
\begin{equation}
\phi_{\omega, {\bf k}_{\perp}} (x_{\parallel}) = 2^{\nu} C e^{i
k_{\parallel} (\infty) x_{\parallel}}. \label{s asym1}
\end{equation}
It describes a wave function after tunnelling (an incoming
anti-particle). In the limit of $x_{\parallel} \ll - L$, we may
use another form for Eq. (\ref{s wave})
\begin{eqnarray}
\phi_{\omega, {\bf k}_{\perp}} (x_{\parallel}) = C e^{- \mu
\frac{x_{\parallel}}{L}} {\rm sech}^{\nu}
(\frac{x_{\parallel}}{L}) \Biggl[\frac{\Gamma(\gamma)
\Gamma(\gamma - \alpha - \beta)}{\Gamma(\gamma - \alpha)
\Gamma(\gamma - \beta)} F (\alpha, \beta; \gamma; 1- \zeta)
\nonumber\\ + \frac{\Gamma(\gamma) \Gamma(\alpha + \beta - \gamma
)}{\Gamma(\alpha) \Gamma(\beta)} (1 - \zeta)^{\gamma - \alpha -
\beta} F (\alpha, \beta; \gamma; 1- \zeta)
 \Biggr]. \label{s anoth}
\end{eqnarray}
In this limit the first term of Eq. (\ref{s anoth}) describes an
incident wave (an incoming particle) having the asymptotic form
\begin{equation}
\phi_{\omega, {\bf k}_{\perp}} (x_{\parallel}) = 2^{\nu} C
\frac{\Gamma(\gamma) \Gamma(\gamma - \alpha -
\beta)}{\Gamma(\gamma - \alpha) \Gamma(\gamma - \beta)} e^{i
k_{\parallel} (- \infty) x_{\parallel}}. \label{s asym2}
\end{equation}
Therefore, from Eqs. (\ref{s asym1}) and (\ref{s asym2}) we can
find the probability for tunnelling
\begin{eqnarray}
P^{\rm b.t}_{{\rm k}_{\perp}} &=& \frac{k_{\parallel} (-
\infty)}{k_{\parallel} (\infty)} \Biggl|\frac{\Gamma(\gamma -
\alpha) \Gamma(\gamma - \beta)}{\Gamma(\gamma) \Gamma(\gamma -
\alpha - \beta)} \Biggr|^2 \nonumber\\ &=& \frac{\sinh \pi \Bigl(L
k_{\parallel} (\infty)\Bigr) \sinh \pi \Bigl(L k_{\parallel}
(-\infty)\Bigr)}{\cosh \pi \Bigl(\frac{L}{2} (k_{\parallel}
(\infty) + k_{\parallel} (- \infty) ) + Q\Bigr) \cosh \pi
\Bigl(\frac{L}{2} (k_{\parallel} (\infty)+ k_{\parallel} (-
\infty)) - Q \Bigr)},
\end{eqnarray}
where $Q = \sqrt{(q E_0 L^2)^2 - 1/4}$. In the large $L$ limit we
obtain approximately the probability for tunnelling
\begin{equation}
P^{\rm b. ~tun.}_{{\rm k}_{\perp}} = \frac{1}{1 + e^{2 S_{{\bf
k}_{\perp}}}}.
\end{equation}
Here, we used the binomial expansion
\begin{eqnarray}
k_{\parallel} (\pm \infty) &=& (qE_0L \pm \omega) \Biggl[ 1 -
\frac{m^2 + {\bf k}_{\perp}^2}{(qE_0L)^2 \Bigl(1 \pm
\frac{\omega}{(qE_0L)^2}\Bigr)^2} \Biggr]^{1/2} \nonumber\\ &=&
qE_0 L \pm \omega - \frac{m^2 + {\bf k}_{\perp}^2}{2qE_0L} \Biggl[
1 \mp \frac{\omega}{qE_0L} + \frac{\omega^2}{(qE_0L)^2}\Biggr]
\nonumber\\ && \quad - \frac{(m^2 + {\bf
k}_{\perp}^2)^2}{8(qE_0L)^3} \Biggl[ 1 \mp \frac{3\omega}{qE_0L} +
\frac{6\omega^2}{(qE_0L)^2}\Biggr] + \cdots .
\end{eqnarray}

Therefore, using instanton action (\ref{s inst}) we are able to
obtain the pair-production rate for bosons and fermions according
to Eqs. (\ref{inh bos rate}) and (\ref{inh fer rate}). Thus we
have shown that in the space-dependent gauge the instanton
interpretation for wave function gives correctly the
pair-production rates for bosons and fermions for two exactly
solvable models.

\section{Magnetic Fields}

A static uniform magnetic field leads only to a real effective
action and thus implies no-pair production \cite{sch}. Recently it
has also been shown that any static magnetic field, having no
imaginary part, does not lead to the pair-production
\cite{dun,sri,dun2}. On the other hand, in canonical method, each
mode of the Klein-Gordon or Dirac equation has a nonzero
reflection probability for static magnetic fields. This issue has
been raised and discussed to interpret the reflection probability
as the pair-production by static localized magnetic fields in Ref.
\cite{sri}. In this section we resolve this issue from the view
point of the instanton interpretation.

Let us consider a static magnetic field in a 4-dimensional
spacetime with the gauge potential
\begin{equation}
A_{\mu} (t, {\bf x}) = (0, A_1 (x_2), 0, 0).
\end{equation}
The magnetic field is given by ${\bf B} = (d A_1 (x_2)/dx_2)
\hat{\bf x}_3$. The Klein-Gordon equation has the form
\begin{equation}
\Biggl[ \frac{\partial^2}{\partial t^2} -
\Biggl(\frac{\partial}{\partial x_1} + i q A_1 (x_2) \Biggr)^2 -
\frac{\partial^2}{\partial x_2^2} - \frac{\partial^2}{\partial
x_3^2} + m^2 \Biggr] \Phi (t, {\bf x}) = 0. \label{kg eq4}
\end{equation}
As in the case of the electric field, each mode of the field
\begin{equation}
\Phi (t, {\bf x}) = e^{i (k_1 x_1 + k_3 z_3 - \omega
t)}\phi_{\omega, k_1, k_3} (x_2)
\end{equation}
leads to a Schr\"{o}dinger-like equation
\begin{equation}
\Biggl[- \frac{1}{2} \frac{d^2}{dx_2^2} + \frac{1}{2}\Biggl(k_1 -
q A_1 ( x_2)\Biggr)^2 \Biggr] \phi_{\omega, k_1, k_3} (x_2) =
\frac{1}{2}(\omega^2 - m^2 - k_3^2) \phi_{\omega, k_1, k_3} (x_2).
\label{mag mod eq}
\end{equation}
As a one-dimensional quantum system, Eq. (\ref{mag mod eq}) has
the potential $(k_1 - q A_1 ( x_2))^2/2$ and the energy $(\omega^2
- m^2 - k_3^2)/2$.

In the case of a uniform magnetic field, the gauge potential $A_1
(x_2) = - B_0 x_2$ is indefinitely unbounded at $x_2 = \pm
\infty$. Then the potential of Eq. (\ref{mag mod eq}) is exactly
that of a harmonic oscillator and the energy is quantized
\begin{equation}
\epsilon_n = q B_0 (2n+1).
\end{equation}
The quantized energy has been used to calculate the effective
action in the uniform magnetic field \cite{hei}. From the view
point of instanton interpretation, there is no pair-production
since there are no finite instantons at all. All would-be
instantons from one spatial infinity to another are infinite and
do not contribute to the tunnelling probability. This result
agrees with that obtained from the proper time and other methods.

We now consider a static localized magnetic field ${\bf B} (x_2) =
B_0 {\rm sech}^2 (x_2/L ) \hat{\bf x}_3$. The gauge potential is
given by $A_1 (y) = - B_0 L \tanh (x_2/L)$. The gauge potential in
Eq. (\ref{mag mod eq}) has two asymptotic values $(k_1 \pm q B_0
L)^2/2$ at $x_2 = \pm \infty$, respectively, and a minimum value
in-between. There is a potential well instead of a potential
barrier, so the reflection probability may be nonzero, though the
tunnelling probability via instantons is zero. Therefore,
according to the instanton interpretation, there is no
pair-production. The instanton interpretation thus resolves the
contradiction between the effective action and the canonical
method raised in Ref. \cite{sri}. Our result also agrees with that
by Dunne and Hall, who showed that the imaginary part of the
effective action vanishes for the static magnetic field considered
above and therefore neither bosons nor fermions are produced in
pairs \cite{dun2,dun3}.

\section{Discussion and Conclusion}

In this paper we have studied the pair-production of bosons and
fermions by static uniform or inhomogeneous electric field. For
these fields we used the space-dependent (Coulomb) gauge and
solved the Klein-Gordon equation. For strong electric fields the
mode-decomposed Klein-Gordon equations have potential barriers
from the gauge potential. The set of wave functions describing
pair-production in quantum field theory is the same of the
standard scattering problem through potential barriers in quantum
mechanics \cite{pad,pad2,han}, in contract with the time-dependent
gauge. Together with the fact that the most dominant contribution
to pair-production rate is the single instanton
\cite{bre,itz,neu,neu2,neu3}, we further propose that all
multi-instantons contribute to the pair-production and
anti-multi-instantons to the annihilation of created pairs and
that the total tunnelling probability from all multi-instantons
and anti-multi-instantons is related with pair-production and the
probability for the vacuum-to-vacuum transition is the probability
for no-pair production. Based on this we derived the
pair-production formula for bosons (\ref{b-p}) and fermions
(\ref{f-p}).

This instanton interpretation means that the single-instanton is
related in a certain way with the single-pair production,
multi-instantons with the multi-pair production and
anti-multi-instantons with the annihilation of the created pairs.
In fact, when the instanton action is large, the single-instanton
is the dominant contribution to one-pair production and
multi-instantons are the dominant contribution to the multi-pair
production. Also it implies the no-pair production when there is
not any tunnelling instanton. In the case of a uniform electric
field, when all the contributions from multi-instantons and
anti-instantons are taken into account, the pair-production rate
for bosons calculated according to the instanton interpretation
recovers the well-known result from the proper time method. By
taking the Pauli-blocking into account the pair-production rate
for fermions is found also to agree with the standard result.
Further the pair-production formula from instanton action yields
the correct forms (\ref{s b-p}) and (\ref{s f-p}) for extremely
strong electric fields, confirming the consistency of the formula
with other methods. Using the instantons obtained in the WKB
(adiabatic) approximation we are also able to provide the formula
for the pair-production rate of bosons and fermions by
inhomogeneous electric fields.

As a by-product we are able to show that any static (localized)
magnetic field cannot produce pairs of bosons or fermions. In the
case of magnetic fields the space-dependent gauge reduces the
Klein-Gordon equation to time-independent Schr\"{o}dinger
equations with potential wells instead of potential barriers of
the electric field case. Since potential wells cannot have finite
instantons and possible infinite instantons from either side of
potential wells give the zero probability for pair-production, the
pair-production of bosons or fermions cannot proceed. As the
nonzero transmission probability through potential wells is the
result without any finite instanton, the instanton interpretation
excludes the possibility of pair-production by a static localized
magnetic field in the canonical method raised in Ref. \cite{sri}.
Therefore we conclude that any static magnetic field does not lead
to pair-production and the canonical method equipped with the
instanton interpretation is compatible with the effective action
method \cite{dun,sri,dun2}.

\acknowledgements

We would like to thank F. C. Khanna and L. Sriramkumar for many
useful discussions, G. Dunne for comments on no-pair production by
a static magnetic field and H. Neuberger for useful information.
S.P.K. would like to express his appreciation for the warm
hospitality of the Theoretical Physics Institute, University of
Alberta. The work of S.P.K. was supported by the Korea Science and
Engineering Foundation under Grant No. 1999-2-112-003-5 and the
work of  D.N.P. by the National Sciences and Engineering Research
Council of Canada.

\end{document}